\documentstyle[prb,aps]{revtex} \def\narrowtext{} \tighten \twocolumn
 
\begin{document}
\draft
 
\title{Temperature-induced spectral weight transfer in 
Bi$_2$Sr$_2$CaCu$_2$O$_{8+\delta}$: a conventional view}
\author{
        J. C. Campuzano,$^{1,2}$
        H. Ding,$^{3}$
        H. Fretwell,$^1$
        J. Mesot,$^{2}$
        A. Kaminski,$^{1}$
        T. Yokoya,$^4$
        T. Takahashi,$^5$
        T. Mochiku,$^5$ and K. Kadowaki$^6$
       }
\address{
         (1) Department of Physics, University of Illinois at Chicago,
             Chicago, IL 60607\\
         (2) Materials Sciences Division, Argonne National Laboratory,
             Argonne, IL 60439 \\
         (3) Department of Physics, Boston College, Chestnut Hill, MA 
             02467\\
         (4) Department of Physics, Tohoku University, 980 Sendai, Japan\\
         (5) National Research Institute for Metals, Sengen, Tsukuba,
             Ibaraki 305, Japan\\
         (6) Institute of Materials Science, University of Tsukuba, 
             Ibaraki 305, Japan\\
         }

\address{%
\begin{minipage}[t]{6.0in}
\begin{abstract}
In contrast with the recent photoemission 
observations by Shen {\it et al.} of anomalous 
temperature-induced momentum-dependent spectral weight transfer in 
Bi$_2$Sr$_2$CaCu$_2$O$_{8+\delta}$, we find that in the same 
material,  in spite of the unusual spectral lineshape change between 
superconducting and normal states, the integrated spectral weight 
displays minimum variation as a function of temperature, which is 
in agreement with the sum rule of angle-resolved photoemission: 
the integrated spectral weight is proportional to the momentum density, 
n(\bf k).
\typeout{polish abstract}
\end{abstract}
\end{minipage}}

\maketitle
\narrowtext
One of the many ways in which high temperature superconductors differ 
from previously studied materials is that their spectra measured by 
angle-resolved photoemission are strongly temperature 
dependent\cite{OLSON_89,DESSAU,MODE}. This 
dependence has been attributed to strong electron-electron 
interactions\cite{RANDERIA,MODE}, which are the dominant 
scattering mechanism for electrons in these materials. Much 
has been learned from studying the spectral lineshapes.  We have shown 
that, under suitable experimental conditions, 
the intensity measured in an angle-resolved photoemission (ARPES) 
experiment is given by \cite{RANDERIA}
\begin{equation}\label{intensity}
I({\bf k},\omega) = I_0({\bf k}) f(\omega) A({\bf k},\omega),
\end{equation}
where ${\bf k}$, the in-plane momentum, gives the location in the 
2D Brillouin zone, and $\omega$ is the energy of the initial state
measured relative to the chemical potential. (Experimentally
$\omega$ is measured relative to the Fermi level of a good metal
like Pt or Au in electrical contact with the sample).
$I_0({\bf k})$ includes all the  kinematical factors and 
the dipole matrix element (squared). It depends, in addition to
${\bf k}$, on the incident photon energy and polarization. $I_0$ 
determines the overall intensity of the measured spectra through the 
dipole selection rules obeyed by the matrix elements, and does not 
affect the spectral lineshape.

The spectral lineshape ($\omega$ dependence) and its $T$ dependence, 
at the low frequencies and temperatures of interest to us, are entirely
controlled by $f(\omega)A({\bf k},\omega)$.
Here $A({\bf k},\omega)$  is the initial state or ``photo-hole'' spectral 
function
$A({\bf k},\omega) = (-1 / \pi){\rm Im}G({\bf k},\omega+i0^+)$, and
the Fermi function $f(\omega) = 1/[\exp(\omega/kT) + 1]$ ensures that
we are only looking at the {\it occupied} part of this
spectral function. Furthermore, the ARPES intensity obeys the sum rule 
such that the integral of a single spectral peak is proportional to 
the momentum distribution, n({\bf k})
\begin{equation}\label{nofk}
\int_{-\infty}^{+\infty}d\omega f(\omega) A({\bf k},\omega) = n({\bf k}).
\end{equation} 
An experimental example from the high-T$_c$ superconductor 
Bi$_2$Sr$_2$CaCu$_2$O$_{8+\delta}$ (Bi2212) can be found in 
Ref.\onlinecite{CAMPUZANO_96}. 

We have shown that, in spite of the unusual spectral lineshape changes 
exhibited by Bi2212 with temperature, the integrated spectral weight 
still satisfies the sum rule (2), such that the integrated weight at, 
or far away from, the Fermi surface (${\bf k}_F$) is temperature 
independent \cite{RANDERIA}. This work has found that as the 
temperature is changed from above to below $T_{c}$ (approximately 90 K), 
lineshape changes occur over an energy scale of order $4\Delta_{0}$, 
where $\Delta_{0}$ is the maximum value of the superconducting gap. 
In general, one does not expect spectral changes to occur over an 
energy range which is orders of magnitude larger than $4\Delta_{0}$, 
unless there is an intervening structural phase transition. Such phase 
transitions have not been observed in clean Bi2212.  

However, a recent photoemission study on Bi2212 by Shen {\it et al.} 
\cite{SHEN} reported anomalous temperature-induced 
momentum-dependent changes between ARPES spectra in the 
superconducting and normal states.  
The reported effect is rather drastic, with spectral changes occurring 
over an energy range as large as the bandwith, $\sim$ 300 meV, 
much larger than the maximum superconducting gap, 
$\Delta_{0}$, ($\sim$30 - 40 meV). 
Also, spectral weight is found to be transferred over a sizable momentum 
vector,~0.45$\pi$, which is much larger than the conventional 
thermal effect, $max(kT, \Delta)/\hbar v_F$, where $v_F$ is the Fermi velocity. 
According to Ref. \cite{SHEN}, this is due to the presence 
of  fluctuating stripes which have charge- and spin-order 
periodicities of (0.45$\pi$, 0).

The topic of spin and charge stripes and their role in high 
temperature superconductors is of current interest\cite{LEVI}. 
There is some evidence that spin and charge stripes exist in 
the high-T$_c$ superconductor $La_{2-x}Sr_{x}CuO_{4}$ doped with Nd. For 
Bi2212, Ref. \cite{SHEN} would represent the 
strongest evidence yet for the presence of charge stripes in this 
material. Thus it is important to rule out other possible explanations 
for such an unusual result. In this paper we wish to point out that 
there are experimental artifacts that can produce the effect observed 
by Shen {\it et al.}, and that these must be first ruled out in order to 
strengthen the case for the observation of stripes by ARPES. 
Here we show that there are two possible effects which can lead to the 
results reported in Ref.\onlinecite{SHEN}, namely a change in emission 
angle in the experiment, and sample aging. After discussing the source 
for these artifacts, we propose experiments designed to check for 
their presence, and in the first case, how to account for it.

Our experiments were performed at the Synchrotron Radiation Center,
Wisconsin, using a high resolution 4-meter normal incidence monochromator.
The high quality single crystals Bi2212, grown by the traveling solvent 
floating zone method, have low defect densities with very long
structural coherence lengths. Also, a crucial requirement is the presence 
of a flat surface after cleaving, as we discuss in detail below. 

We start by showing the normal behavior of the spectral weight 
transfer with temperature. In Fig. 1 we show, as an example, a 
slightly underdoped (T$_c$ = 89K) sample which was a very flat surface 
and was measured from high temperature ($\sim$300K) to low temperature 
(14K) in a short period of time (less than 3 hours) in a good vacuum 
($\sim$5x10$^{-11}$) to ensure the 
absence of surface contamination. Fig. 1a displays spectra along 
(0,0) to $(\pi,0)$ at different temperatures. First, the broad normal 
state peak evolves into a sharp peak, dip and broad peak (hump) in 
the superconducting state. This has been previously described in 
detail\cite{MODE}, and occurs as a result of a rapid variation with
$\omega$ of the electron 
damping in the superconducting state. Second, spectra show 
identical high-binding-energy features (broad peaks and tails) beyond 
binding energies of 4$\Delta_{0}$. No anomalous weight transfer is observed, 
as shown in Fig. 1b, which plots the area under each spectra 
(from -0.6 eV to 0.1 eV) as a function of momentum. The areas of the 
two states are the same for each {\bf k} point within the experimental 
uncertainty. The fact that most experiments, under carefully controlled 
experimental conditions, show spectra which are "well-behaved" 
immediately throws suspicion on unusual results, such as those shown 
in Fig. 2 (and those of Ref. \cite{SHEN}).

In Fig. 2 we show ARPES spectra of a slightly overdoped 
Bi2212 sample (T$_c$ = 87K) along (0,0) to ($\pi$,0), extending to 
the second Brillioun Zone with a high density of {$\bf k$} points. 
Solid curves are for the normal state at 105K, dotted ones are for the 
superconducting state at 13K. It is clear from Fig. 2 that, unlike the 
case of Fig. 1, the spectra are quite different between normal and 
superconducting states. The overall spectral weight changes between 
the two temperatures. The change is most noticeable at small 
momenta, where the dispersion is strongest (see spectra between 
(0.37,0)$\pi/a$ and (0.56,0)$\pi/a$). The self-energy 
is not expected to change significantly at such high binding energies 
with an increase in temperature of only $\sim$100 K. The change is 
quantified in Fig. 2b, which shows a plot of the integrated intensity 
of the data in Fig. 2a. Note that this 
observation is exactly what is described in Ref. \cite{SHEN}. 
We should point out that we use a different normalization 
procedure from that of Ref. \cite{SHEN}. Our data are simply 
normalized to the photon flux by normalizing each spectra to the same 
second order intensity well above the Fermi energy. Therefore our 
spectra are plotted as being proportional to electrons per photon. 
The use of this single normalization procedure, and an artifact to be 
described below, result in discrepancies at high binding energy between 
the two states. The same problem is present in Ref. \cite{SHEN}, but 
there the authors use an additional normalization (different for 
each {\bf k} point) to match the high binding energy tails of the two 
states. This procedure is highly undersirable, since it 
rescales the spectra in unknown ways and is a sign of potential 
experimental problems, as discussed below.

That an artifact is responsible for the unusual spectral weight 
transfer is confirmed by a closer inspection of the data 
in Fig. 2. If, instead of comparing spectra in the normal and 
superconducting states at nominally the same momenta, we 
instead translate the normal state spectra by 
$(0.09\pi/a,0)$ with respect to the superconducting spectra 
(equivalent to a two-degree change of the emission angle along 
k$_{x}$), then not only do most of the broad features 
now agree, but the high-binding-energy tails also match (see Fig. 
3a) -- in agreement with the behavior of most of our data. By plotting the 
integrated spectral weight (from -0.6 eV to 0.1 eV) as a function 
of momentum (Fig. 3b), one can clearly see 
this effect: by a simple shift in k$_x$ of 0.09$\pi$/a, one obtains 
a much better agreement of the integrated weight between the two 
states, akin to the spectra shown in Fig. 1. 

This immediately suggests a possible cause for the experimental 
artifact, a possible experimental check, and a method for correcting 
the data. This particular case of large scale spectral weight 
transfer is caused by sample movement, which necessarily occurs in the 
experiment. As the sample temperature is changed from the normal to 
the superconducting regime, the sample changes position because of 
the thermal expansion of the long cryostat on which the sample is 
mounted \cite{CRYO}. During an experimental run, the sample position 
must be continuously adjusted to account for such movements. 
Such position changes  can shift emission angle, especially when the 
sample surface is not flat. Fortunately, given a sufficiently high 
number of spectra, this effect can be easily corrected for, as 
shown in Fig. 3a 

The second type of effect, sample aging, unfortunately, cannot be easily 
corrected for. Sample aging also produces similar large scale changes in the 
data \cite{UNDERFS}. This effect is more pronounced in a lower vacuum, 
as shown in Fig. 4, where the chamber pressure is $\sim$4x10$^{-10}$, 
almost one order of magnitude worse than the one in Fig. 1. 
Figure 4 shows ARPES spectra at the most dispersive region 
along (0,0) to ($\pi$,0) for a slightly overdoped Bi2212 sample 
(T$_{c}$ = 88K). Dotted curves are for the superconducting state at 50K and 
solid ones correspond to the normal state at 100K. The sample was cleaved 
and measured at 50K, then heated to 100K. Notice that the 100K spectra 
shift considerably towards the Fermi energy, similar to the observation 
in Fig. 2 and Ref. \cite{SHEN}. However, the cause of this is not
emission angle change, rather it is sample aging, which in turn changes the 
doping level on the surface \cite{UNDERFS}. This effect
also has a characteristic signature, namely the shift of more aged
spectra is always towards lower binding energy, unlike in the previous
case, where the direction of shift is determined by temperature.
 
We have measured several samples with various doping levels 
to check for the presence of a weight transfer. 
We find the effects discussed in Fig. 2 and Fig. 4 only when 
the samples have a 
non-flat surface or they visibly age, or when there is 
an instability in the synchrotron beam. Therefore, it is reasonable 
to conclude, at least in our case, that the anomalous weight transfer 
is an experimental artifact. This conclusion strongly suggests that the 
artifacts described here need to be ruled out in the case of the samples 
used by Shen {\it et al} in order to strenghten the case for an 
anomalous spectral weight transfer in Zn-doped samples. 

Several possible experiments suggests themselves to this effect. One 
must first ensure that the sample position remains constant during 
temperature changes, which can easily be accomplished to high 
precision using existing optical techniques, such as theodolites or 
laser interferometers. One must also ensure that the {\it cleaved} 
surfaces are optically flat, which, again, can be accomplished 
{\it in situ} with laser reflectometry. The data can also be checked 
for internal consistency. A plot of the integrated intensity vs momenta, 
such as those shown in Figs 2b and 3b, will indicate the presence of 
an artifact. But most importantly, if the anomalous weight transfer is 
indeed due to the presence of stripes which shift the spectra by a given 
momentum transfer $Q$, one should check that the spectral weight is {\it 
periodic} in $Q$. In our case, as clearly indicated by Fig. 3b, it is 
not periodic, and therefore the effect cannot be ascribed 
to an additional periodicity introduced by stripes. 

This work was supported the National Science Foundation DMR 9624048, 
and DMR 91-20000 through the Science and Technology Center for
Superconductivity, and by the U. S. Dept. of Energy,
Basic Energy Sciences, under contract W-31-109-ENG-38.
The Synchrotron Radiation Center is supported by NSF grant DMR-9212658.

\begin{figure}
\caption{(a). Temperature-dependent ARPES spectra in the (0,0) to 
($\pi$,0) direction from a slightly underdoped Bi2212 with T$_c$ = 89K. 
For each momentum, there are three curves measured at different temperatures, 
dashed curves are from 192K to 141K, solid ones from 137K to 81K, 
thick dotted ones from 76K to 15K. (b). The corresponding integrated 
spectral weight of the spectra in (a).
}
\label{fig1}
\end{figure}

\begin{figure}
\caption{a) ARPES spectra along (0,0) to ($\pi$,0) in the normal state at 
105K (solid curves) and the superconducting state at 13K (dashed 
curves) from a slightly overdoped Bi2212 sample with T$_c$ = 87K; b) 
Plot of the differences of the integrated intensities of the normal 
and superconducting states as a function of angle for the data in (a).
}
\label{fig2}
\end{figure}

\begin{figure}
\caption{(a). Same ARPES data as in Fig. 2, but with the normal state 
shifted by $0.09\pi/a$ in the k$_x$ direction, which corresponds to 
two-degree emission angle change. (b). Integrated areas as a function 
of momentum, k$_x$. Note the discrepancy between normal state areas 
(dotted line with empty circles) and superconducting ones (solid line 
with asterisks) can be corrected by a small shift of normal state 
data (dashed line with filled circles). 
}
\label{fig3}
\end{figure}

\begin{figure}
\caption{ARPES spectra at the most dispersive region along (0,0) to
($\pi$,0) in the normal state at 100K (solid curves) and the 
superconducting state at 50K (dashed curves) from a slightly overdoped 
Bi2212 sample with T$_{c}$=88K.
}
\label{fig4}
\end{figure}


\begin{references}
\bibitem{CRYO}
In both Shen's experiment and ours, the sample is attached to a 
cryostat of length of order a meter, which shortens by more than 4 mm 
as the temperature drops from room to base temperature. This length is 
larger than typical sample diameters, so that continuous sample 
position adjustment is necessary during an experimental run.
\end{references}
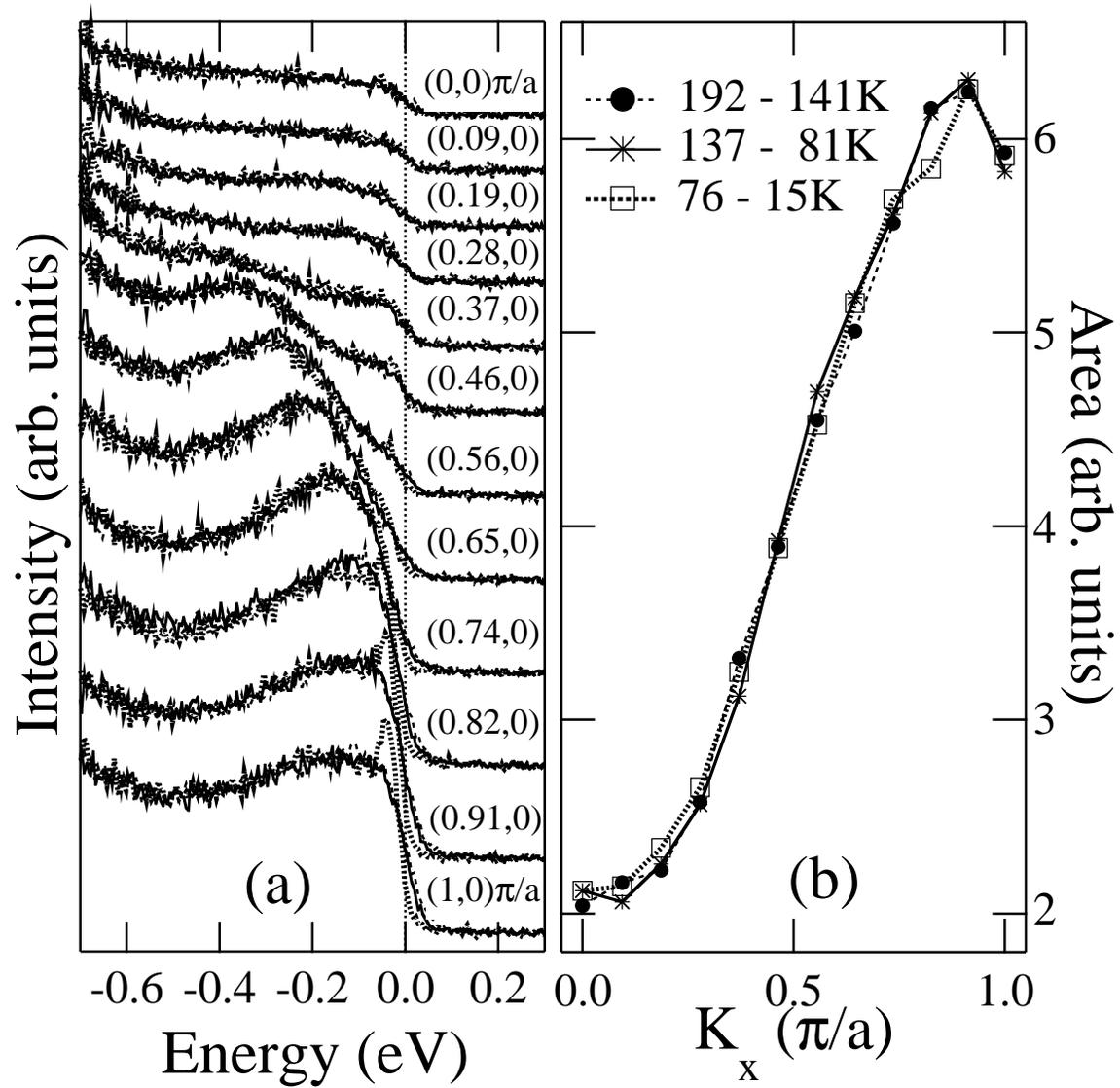
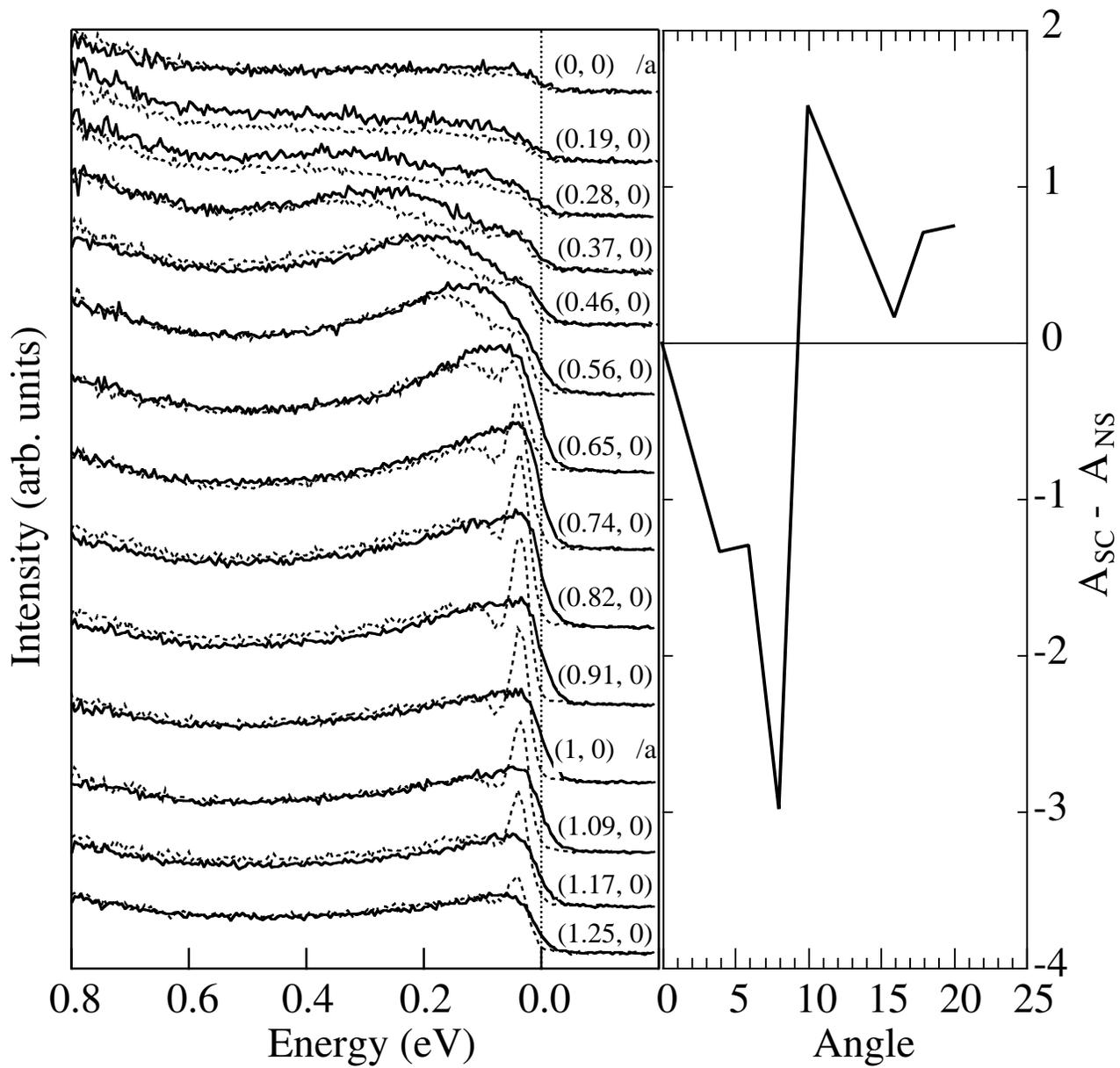
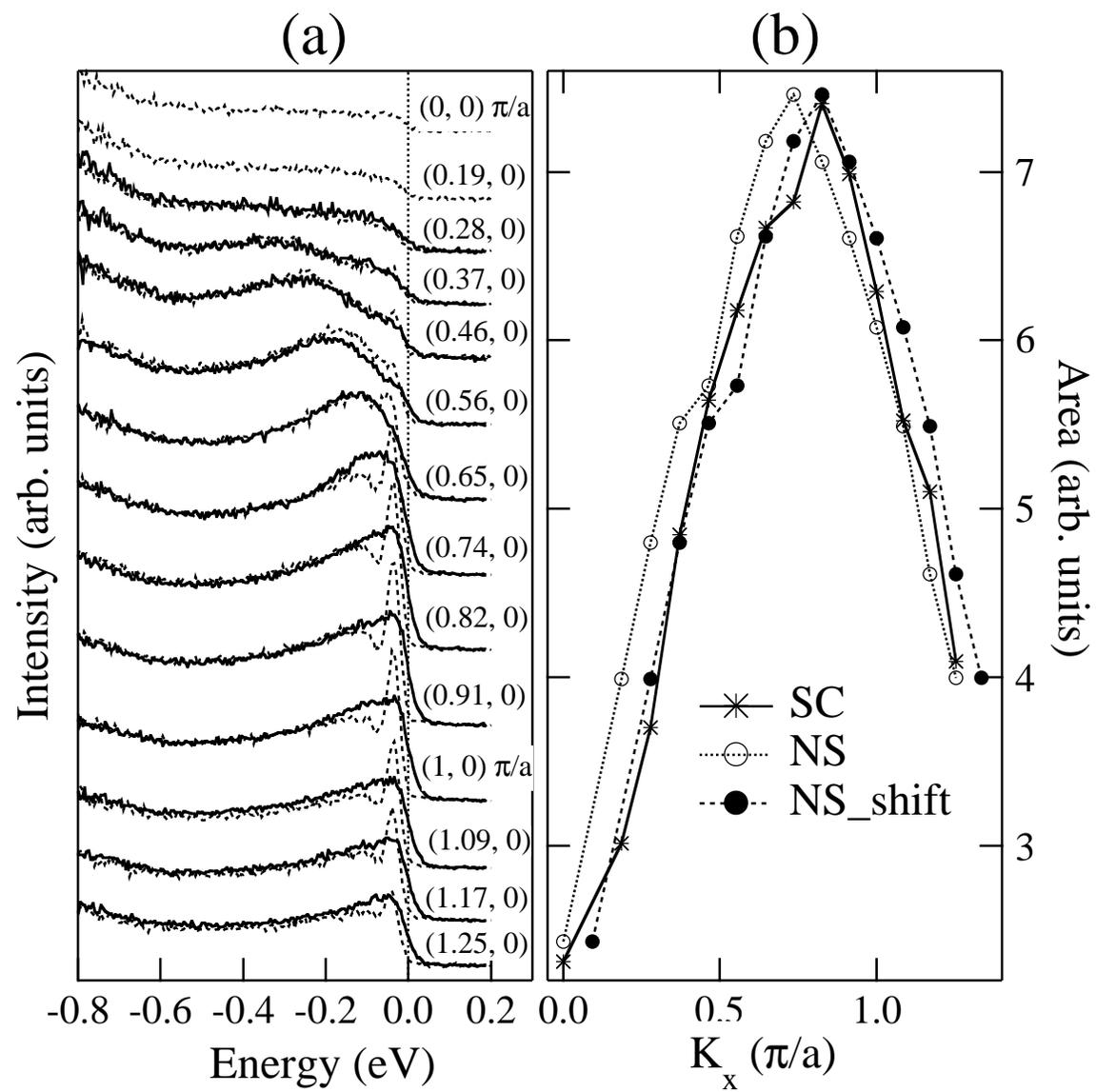
\end{document}